\documentclass[prd,aps,superscriptaddress]{revtex4}

\usepackage{graphicx}
\usepackage{dcolumn}
\usepackage{bm}
\usepackage{latexsym, amsmath, amssymb}

\def\be{\begin{equation}}
\def\ee{\end{equation}}

\begin{document}
\voffset = 0.3 true in
\topmargin = -1.0 true in 

\title{Charmonium-Nucleon Dissociation Cross Sections in the Quark Model}

\author{J.P. Hilbert}
\affiliation{Department of Physics and Astronomy, University of Pittsburgh,
Pittsburgh PA 15260}

\author{N. Black}
\affiliation{Utah Center for Advanced Imaging Research,
University of Utah, Salt Lake City, UT  84108}
\author{T. Barnes}
\affiliation{Physics Division, Oak Ridge National Laboratory,
Oak Ridge, TN 37831}
\affiliation{Department of Physics and Astronomy, University of Tennessee,
Knoxville, TN 37996}

\author{E.S. Swanson}
\affiliation{Department of Physics and Astronomy, University of Pittsburgh,
Pittsburgh PA 15260}

\vskip .5 true cm
\begin{abstract}
Charmonium dissociation cross sections due to flavor-exchange 
charmonium-baryon scattering are computed in the constituent quark model. 
We present results for inelastic 
$J/\psi N$ 
and
$\eta_c N$
scattering amplitudes and cross sections into 
46 final channels, including final states composed of
various combinations of $D$, $D^*$, $\Sigma_c$, and $\Lambda_c$. 
These results are relevant to experimental searches for the deconfined 
phase of quark matter, and may be useful in identifying the contribution
of initial $c\bar c$ production to the open-charm final states observed 
at RHIC through the characteristic flavor ratios of certain channels.
These results are also of interest to possible 
charmonium-nucleon bound states.
\end{abstract}

\maketitle

\section{Introduction}

The production of heavy quarkonium in heavy ion collisions 
has long been considered a possible diagnostic for 
the appearance of exotic QCD phases 
in relativistic heavy ion collisions \cite{ms,vogt}. 
In particular it has been anticipated that the charmonium production
cross section in central $AA$ reactions will be 
suppressed if a quark gluon plasma (QGP) is formed, since the 
long-range $c\bar c$ confining potential will be screened within the QGP.
However, this diagnostic can be confounded by subsequent charmonium 
dissociation due to inelastic hadron rescattering ``comover absorption", 
which also contributes to the depletion of initially produced charmonium. 
Alternatively, charmonium can be 
regenerated due to rescattering of open-charm hadrons 
in the late stages of heavy ion collisions \cite{regen}. 
It is clear that a thorough understanding of these soft hadronic 
final state interactions is required before one can make confident
statements regarding QGP production based on charmonium 
production cross sections \cite{barnes}. The usefulness of 
an understanding of these soft processes is now expected 
to extend well into the deconfined phase, 
since recent lattice computations indicate that low lying charmonia 
survive as resonances up to $T \approx 3T_c$ \cite{latt}.

Unfortunately, little experimental information exists 
regarding these charmonium dissociation cross sections. 
Some simple phenomenological estimates based on 
`pre-vector meson dominance' \cite{hk} and 
absorption cross sections in heavy ion collisions \cite{ghq} give 
$\sigma_{tot}^{\psi N} \propto s^{0.22}$ mb and 
$\sigma_{tot}^{\psi N}(\sqrt{s} \approx 10$~GeV$) \approx 6$~mb respectively.
Alternatively, $J/\psi$ photoproduction yields a $J/\psi\, N$ cross section of
approximately 3.5 mb\cite{RLA} while a combined analysis of  $J/\psi$ production from
$p+A$ collisions gives a result of 7 mb \cite{K}.
For more detailed predictions of the relevant near-threshold cross sections
one must employ theoretical models of these scattering processes.

Theoretical estimates have employed a variety of methods, and (perhaps 
not surprisingly in view of the lack of low-energy experimental data) 
predict cross sections that vary over several orders of magnitude in the 
relevant kinematic regime. Early estimates by Kharzeev and Satz 
\cite{ks} using the color-dipole diffractive model of Bhanot and Peskin 
\cite{BP} (which is only justified at high energies)
gave extremely small near-threshold $J/\psi N$ total cross sections, 
typically of microbarn scale. More recently, $J/\psi N$ dissociation 
cross sections have been estimated using meson exchange models
\cite{mex,haglin}, assuming for example $t$-channel charmed meson exchange 
or an SU(4)-symmetric hadron effective lagrangian. Results 
from these models for meson-$J/\psi$ scattering cross sections 
near thresholds are typically in the few mb range. Similar 
computations have been reported for $J/\psi N$ scattering
\cite{haglin,mex2}, which also find total cross sections near threshold  
in the few mb scale. Although these meson exchange models are of great interest 
as possibly realistic descriptions of these near-threshold processes
they suffer from uncertainties due to 
poorly understood vertex form factors, Fock space truncations in the 
set of exchanged particles, and the questionable assumption of higher 
symmetry groups such as SU(4) (this implicitly assumes a close 
relationship between the dynamics of pions and heavy-quark $c\bar c$ mesons). 
Regarding hadronic form factors, several groups \cite{mexFF} have found 
a strong suppression of the predicted dissociation cross sections 
on incorporating plausible hadronic form factors.

Many of the problems encountered in previous approaches are avoided
if one implements a ``microscopic" quark-gluon description of
these scattering processes, for example using the constituent quark model.
The earliest application of this approach to charmonium dissociation is
the work of Martin {\it et al.}\cite{mbq}, who applied the method of 
Ref.\cite{Bar92} to $J/\psi \pi \to D^*\bar D$ and 
$D^*\bar D^*$ scattering. Martin {\it et al.} assumed that 
the confining interaction only operated between $q$-$\bar q$ pairs;
a more conventional quark-gluon model of charmonium dissociation
cross sections based on the usual $\lambda \cdot \lambda$ 
color structure has been developed by Wong {\it et al.} \cite{wsb}. 
These approaches describe hadronic interactions and
bound states in terms of the nonrelativistic quark model, and usually assume 
that the scattering amplitudes are given to sufficient accuracy at Born order 
in the quark-gluon interaction (this can be relaxed of course). 
At Born order in these valence $q\bar q$ annihilation free channels
hadron-hadron scattering occurs by constituent interchange, and the scattering
amplitudes and cross sections can be derived analytically if sufficiently
simple wavefunctions are employed (for example simple harmonic oscillator
forms). Applications of this approach to a wide range of similar 
scattering processes without valence annihilation, such as I=2 $\pi\pi$ 
\cite{Bar92,qq-BS}, $KN$ \cite{KN-BS} and $NN$ \cite{NN-BS} 
have shown that this leads to numerically realistic results 
for many short-ranged S-wave scattering processes. 

In this paper we report the first results for charmonium dissociation 
due to scattering from nucleons in this type of constituent quark
scattering model \cite{m}. These amplitudes are also relevant 
to the time-reversed process of charmonium regeneration, and may 
prove useful in future studies of bound states of charm and nuclei, 
as suggested by Brodsky {\it et al.} \cite{Brodsky}.

\section{Quark Model Scattering Formalism}
\label{formSect}

The Born-order quark interchange model approximates hadron-hadron 
scattering as due to a single interaction of the standard quark-model 
interaction Hamiltonian $H_I$ between all constituent pairs in different
hadrons \cite{Bar92}. For the calculations reported here we 
employ a constituent quark model interaction of the form 

\begin{equation}
H_I = \sum_{ij}\left( \frac{\alpha_s}{r_{ij}} - \frac{3}{4}br_{ij} 
- \frac{8 \alpha_s \sigma^3}{3 \sqrt{\pi} m_i m_j}
{\rm e}^{-\sigma^2 r_{ij}^2}\, S_i\cdot S_j \right) T_i\cdot T_j
\label{HI}
\end{equation}
where the sum extends over all quarks and antiquarks. 
The three terms in this expression are respectively the color Coulomb interaction, 
linear confinement, and a regularized contact spin-spin hyperfine 
interaction. The model parameters have elsewhere been fitted to  
meson and baryon spectra (see Appendix~\ref{app1} for their values).

The color structure of Eq. \ref{HI} is given by the usual quadratic
perturbative form $T\cdot T$, where $\vec T = \vec\lambda/2$ 
for a quark and $-\vec\lambda^*/2$ for an antiquark ($\lambda$ is a 
Gell-Mann matrix). Meson-baryon systems do not have trivial color 
dynamics ({\it i.e.}, a fixed color state with an overall constant 
color factor), in contrast to mesons and baryons individually, so the assumed 
color structure of Eq.\ref{HI} affects the relative amplitudes for different 
channels. Of course this $T\cdot T$ color structure 
represents a severe truncation of the full dynamics of the gluonic degrees
of freedom. This model does however reproduce the relevant low-energy 
features of more complete models, and the $T\cdot T$ form  
is known to be realistic in describing lattice confinement potentials 
as well as the interactions of low lying hadrons.

We now consider the generic inelastic two-body scattering process 
$AB \to CD$, where $A$ is a $c\bar c$ charmonium state, $B$ is a nucleon,
$C$ is an open-charm $(n\bar c)$ meson, and $D$ is an open-charm 
$(nnc)$ baryon. (Here, $n$ is a light $u$ or $d$ quark.) 
It is convenient to label the quarks as 
$[\bar c c]_{12} [q_1 q_2 q_3]_{345}$. We choose the (45) quark pair to have 
definite symmetry under  $P_{45}$ quark interchange.
Quark interchange scattering in all the systems considered
here involves the specific quark permutations 
$P_{23}$ and $P_{23}P_{34}= P_{34}P_{24}$.  
By inspection, $P_{23}$ permutations give rise to 
$D^{-(*)} \Sigma_c^{++}$ final states in $J/\psi(\eta_c)p$ collisions, 
and $\bar D^{0(*)} \Lambda_c^0(\Sigma_c^0)$ states in
$J/\psi(\eta_c)n$ collisions. 
Similarly, $J/\psi(\eta_c)$ interactions involving the permutation operator
$P_{23}P_{34}$ give $\bar D^{0(*)} \Lambda_c^+(\Sigma_c^+)$ final states, and 
from $J/\psi(\eta_c)n$ one produces $D^{-(*)} \Lambda_c^+(\Sigma_c^+)$ 
final states.

Due to $P_{45}$ symmetry only four unique spatial matrix 
elements are encountered; these are summarized in Table~\ref{SymTab}. 
The first row gives labels specifying the interaction potentials $V_{ij}$.
Thus for example $\bar d_1$ represents a spatial integral of
the form $\int \psi^*_{13} \psi^*_{245} V_{13} \psi_{12} \psi_{345}$.

\begin{table}[h]
\caption{Spatial Symmetries}
\label{SymTab}
\begin{tabular}{l|llllll}
perm & 13 & 14 & 15 & 23 & 24 & 25 \\
\hline
$P_{23}$ & ${\bar d_1}^{\phantom{X}}$ & $\bar d_2$ & $\bar d_2$ & $d_1$ & $d_2$ & $d_2$ \\
$P_{23}P_{34}$ & $\bar d_2$ & $\bar d_1$ & $\bar d_2$ & $d_2$ & $d_1$ & $d_2$ \\
\end{tabular}
\end{table}

\begin{figure}
\includegraphics[scale=0.6]{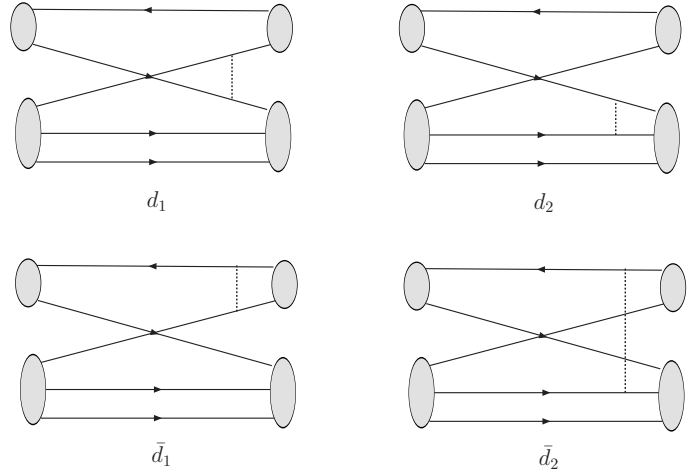}
\caption{The four quark-interchange meson-baryon scattering diagrams
in the prior formalism.}
\label{qex}
\end{figure}

The four unique matrix elements correspond to the four diagrams shown 
in Fig.\ref{qex}. Note that quark line rearrangement is required to give
a nonzero Born-order scattering amplitude, due to the color structure 
assumed for the quark model interaction. We follow the procedures described 
in Ref.\cite{Bar92} in evaluating these diagrams. We assume the `prior' 
form of the scattering amplitude here, in which the separation of the 
full Hamiltonian into free and interaction parts is specified 
by the initial hadrons.  

Hadronic wavefunctions may generically be written as linear combinations of
product basis vectors of the form $\Psi = {\cal C} \chi \Xi \Phi$,
where the individual factors are the color, spin, flavor, and 
spatial wavefunctions respectively. Explicit wavefunctions for the mesons
and baryons considered here are given in Appendix~\ref{app1}.
Because of this factorizability of the hadron wavefunctions,
which involve single factored terms in the cases we consider, the
contribution of each quark diagram to the full scattering amplitude 
can be written as the sum of products of individual color, spin, flavor,
and spatial matrix elements. 

The Born order scattering amplitudes are then constructed by summing 
matrix elements of the interaction potential.
For the case of $P_{23}$ scattering, for example, we have

\begin{eqnarray}
{\cal A} &=& 
  -\langle P_{23} \Xi_C \Xi_D| \Xi_A \Xi_B\rangle \cdot
 \sum_{(ij)} \Bigg[ \langle P_{23} {\cal C}_C {\cal C}_D | T_i \cdot T_j | {\cal C}_A {\cal C}_B\rangle \cdot \nonumber \\
  && \langle P_{23} \chi_C \chi_D| S_i\cdot S_j | \chi_A \chi_B \rangle  \cdot
  \langle P_{23} \psi_C \psi_D | V_{ij} | \psi_A \psi_B \rangle \Bigg].
\end{eqnarray}
The spin factor $S_i\cdot S_j$ 
is replaced by unity when considering central (pure potential) interactions.
This expression can be simplified using the symmetries and notation discussed 
above to obtain
${\cal A} = - \vec w \cdot \vec d$,
where $\vec d = (d_1, d_2, \bar d_1, \bar d_2)$ and $\vec w$ is a weight 
vector arising from flavor, color, and spin matrix elements. 
The weight vectors used in this work are listed in Appendix~\ref{wApp}.

Finally, the charmonium dissociation cross sections were computed from these 
amplitudes using the expression

\begin{equation}
\sigma = \frac{\mu_{AB} \mu_{CD}}{4 \pi^2} 
\frac{k_f}{k_i} \int |{\cal A}|^2 d \Omega
\end{equation}
where $\mu_{AB} = E_A E_B/(E_A+E_B)$.  
Charmonium regeneration cross sections can be also obtained using this 
formalism, since they are related to the dissociation processes by 
time reversal.

\section{Charmonium Dissociation Cross Sections}

A total of 46 exclusive charmonium-nucleon inelastic 
scattering processes were considered, 
involving nucleons and the $J/\psi$ or $\eta_c$ in the initial state 
and all kinematically accessible open-charm 
S-wave-meson--baryon channels in the final state. Specifically, we have computed 
total cross sections for the following reactions:

\begin{equation}
J/\psi p; \eta_c p \to \bar D^0 \Lambda_c^+; \bar D^0 \Sigma_c^+; 
\bar D^{0*}\Lambda_c^+; \bar D^{0*}\Sigma_c^+; D^-\Sigma_c^{++}; 
D^{-*}\Sigma_c^{++}
\end{equation}

\begin{equation}
J/\psi n; \eta_c n \to \bar D^0 \Sigma_c^0; \bar D^{0*}\Sigma_c^0; 
D^{-}\Sigma_c^+; D^{-*}\Sigma_c^{+}; D^{-}\Lambda_c^{+}; 
D^{-*}\Lambda_c^+
\end{equation}

\begin{equation}
[J/\psi p]_{3/2} \to \bar D^{0*}\Sigma_{c\, 3/2}^+; 
D^{-*}\Sigma_{c\, 3/2}^{++}; \bar D^{0}\Sigma_{c\, 3/2}^+; 
D^{-}\Sigma_{c\, 3/2}^{++}
\end{equation}

\begin{equation}
[J/\psi n]_{3/2} \to \bar D^{0*}\Sigma_{c\, 3/2}^0; 
D^{-*}\Sigma_{c\, 3/2}^{+}; \bar D^{0}\Sigma_{c\, 3/2}^0; 
D^{-}\Sigma_{c\, 3/2}^{+}
\end{equation}

\begin{equation}
[J/\psi p]_{1/2}; \eta_c p \to \bar D^{0*} \Sigma_{c\, 3/2}^+; 
D^{-*} \Sigma_{c\, 3/2}^{++}
\end{equation}
and

\begin{equation}
[J/\psi n]_{1/2}; \eta_c n \to  \bar D^{0*} \Sigma_{c\, 3/2}^0; 
D^{-*} \Sigma_{c\, 3/2}^{+}.
\end{equation}

Isospin symmetry and the equivalence of many quark line diagrams 
imply that all the $J/\psi(\eta_c)n$ amplitudes are simply related to  
$J/\psi(\eta_c)p$ amplitudes, as follows:

\be
{\cal A}(J/\psi(\eta_c) n \to \bar D^{0(*)}\Sigma_c^0) = 
{\cal A}(J/\psi(\eta_c)p \to D^{-(*)}\Sigma_c^{++})
\ee
and
\be
{\cal A}(J/\psi(\eta_c) n \to D^{-(*)}\Sigma_c^+(\Lambda_c^+)) = 
{\cal A}(J/\psi(\eta_c)p \to \bar D^{0(*)}\Sigma_c^{+}(\Lambda_c^{+})).
\ee
Thus 23 unique amplitudes remain to be computed.

An additional isospin relation between these 
reactions can be derived because the 
$D^- \Sigma_c^{++}$ and $\bar D^0\Sigma_c^+$ states both couple to 
$|II_z\rangle = |\frac{1}{2} \frac{1}{2}\rangle$. Thus 

\be
\frac{{\cal A}(J/\psi p \to \bar D^0 \Sigma_c^+)}
{{\cal A}(J/\psi p \to D^- \Sigma_c^{++})} 
= -\frac{1}{\sqrt{2}}.
\ee
This relationship evidently holds for the weights reported in Appendix 
\ref{wApp}, which were derived without making this assumption.

\begin{figure}[h]
\includegraphics[width=7cm,angle=0]{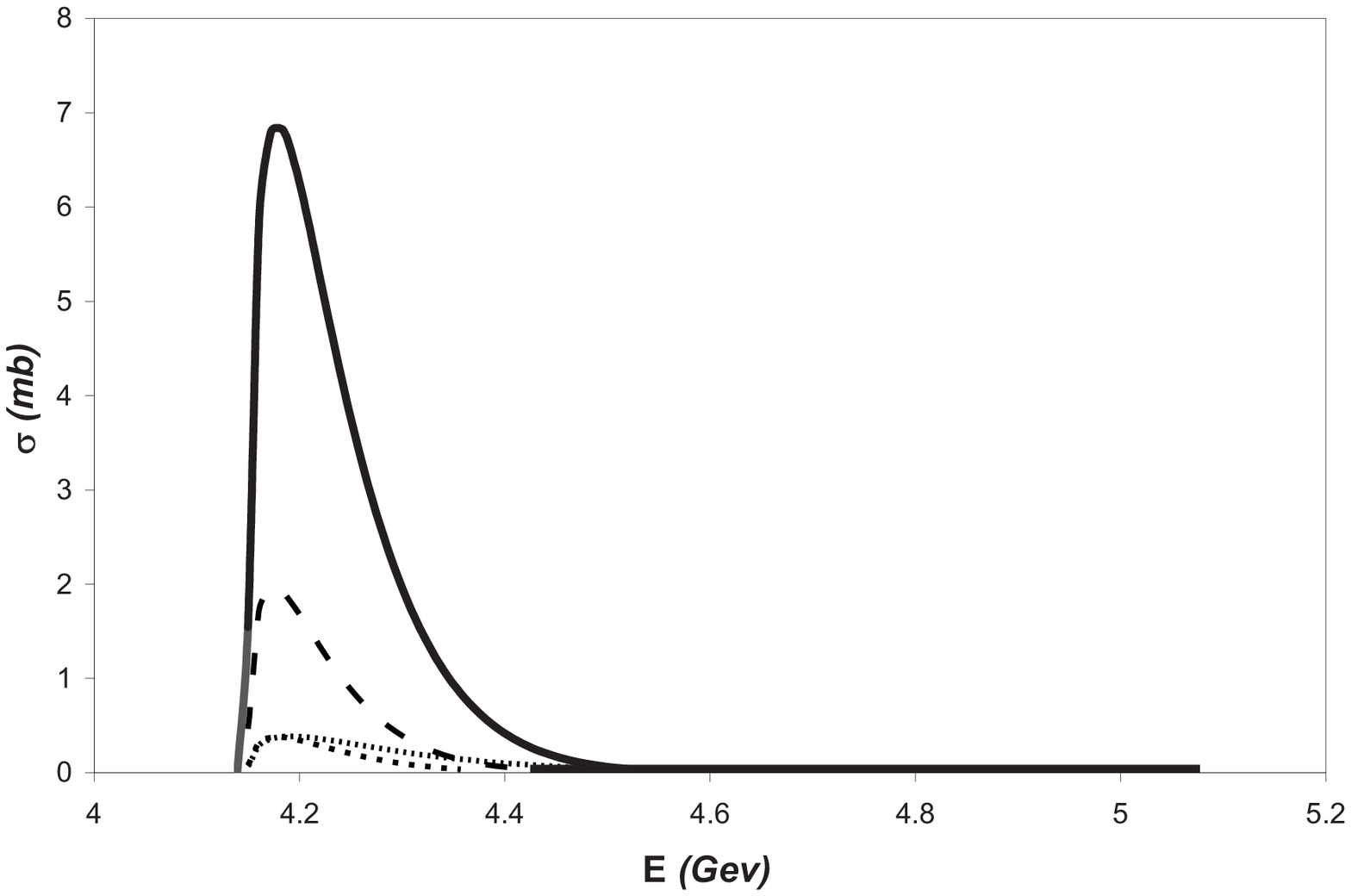}
\hskip 1 true cm
\includegraphics[width=7cm,angle=0]{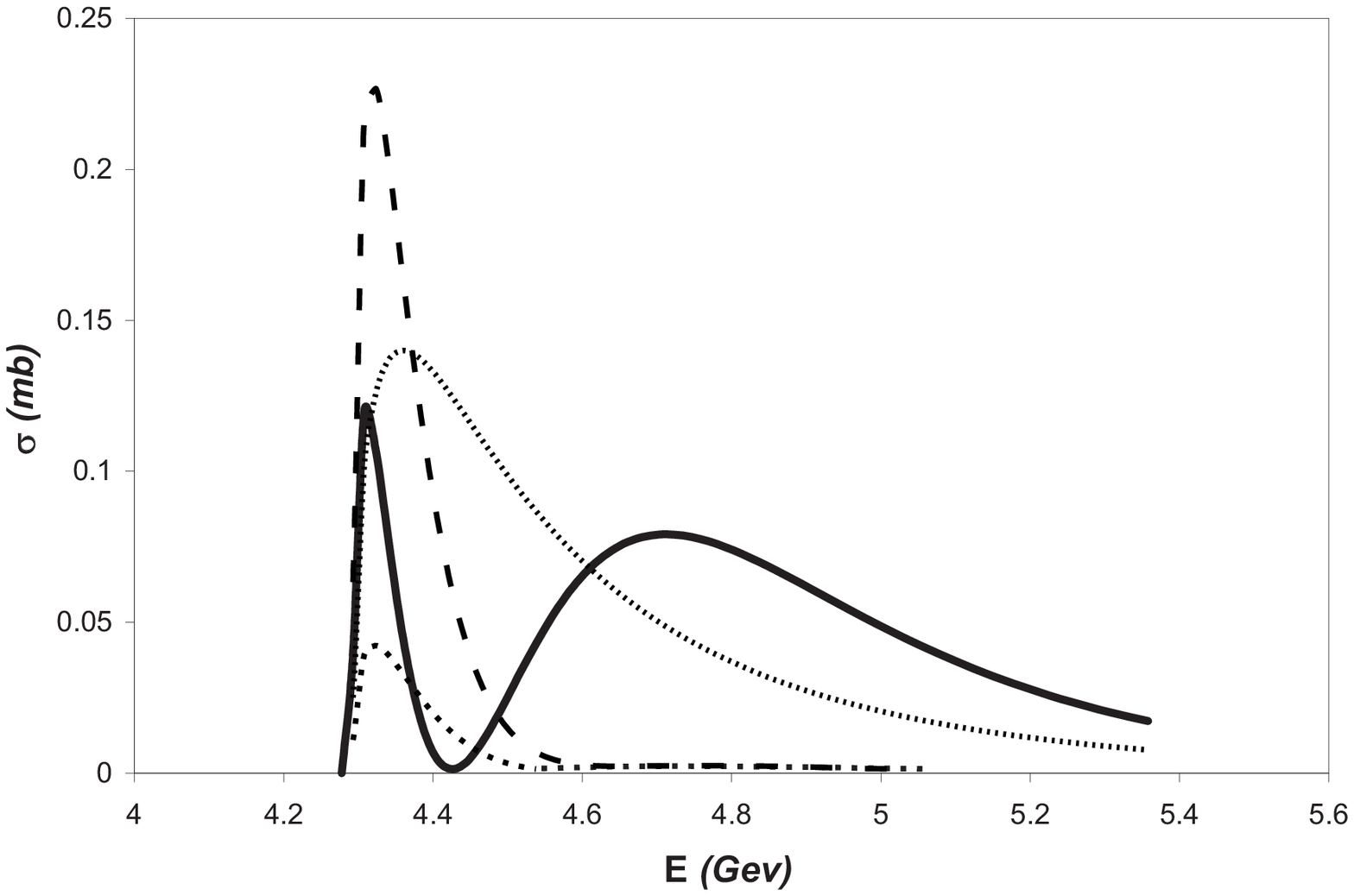}
\caption{$J/\psi p \to \bar D^0 \Lambda_c^+$ (left) $J/\psi p \to \bar D^{0*}\Lambda_c^+$ (right).
Curves are: total cross section (solid), hyperfine (dotted), linear (dashed), Coulomb (small dash).}
\label{xsecFig}
\end{figure}

\begin{figure}[h]
\includegraphics[width=7cm,angle=0]{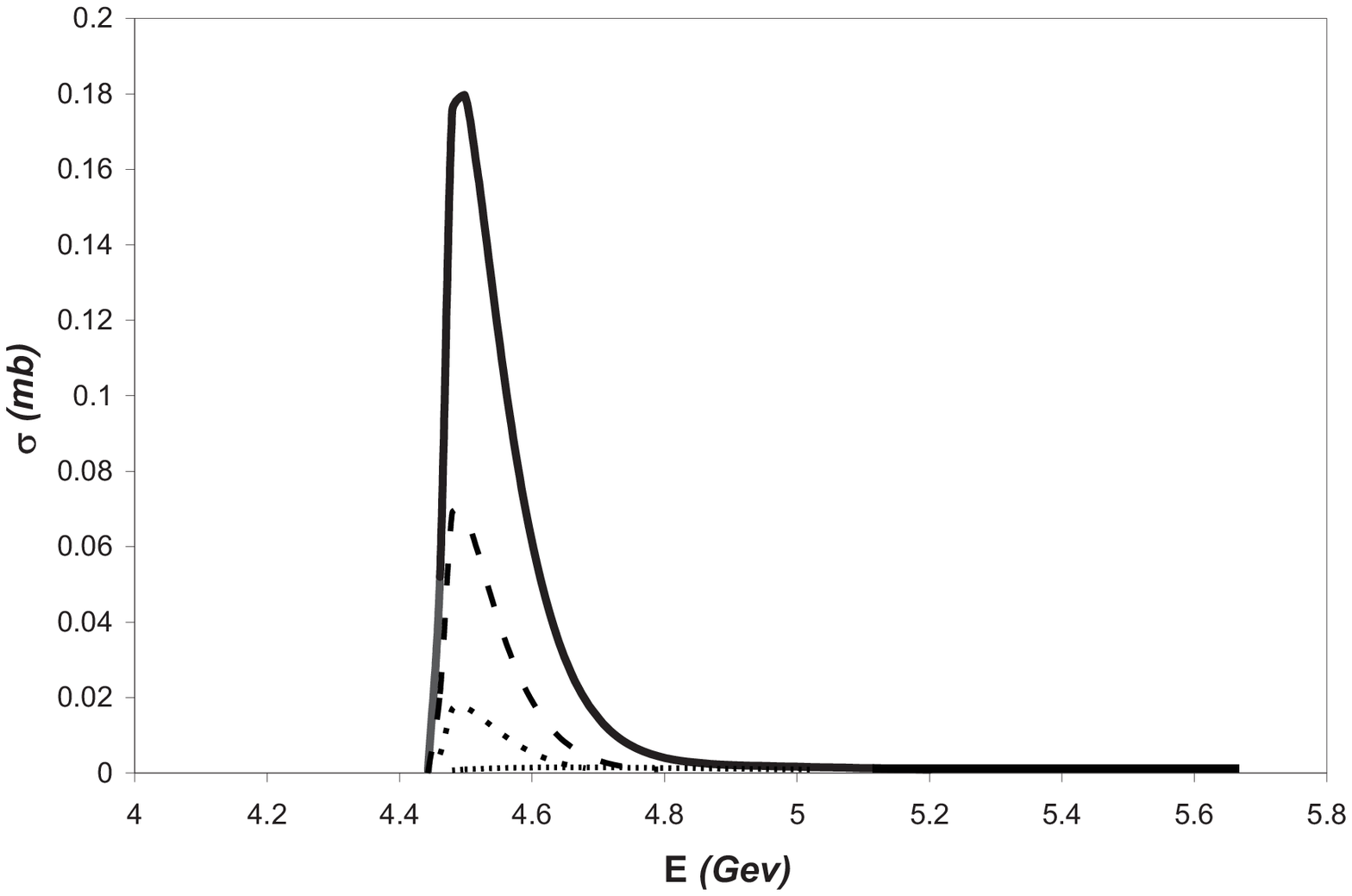}
\hskip 1 true cm
\includegraphics[width=7cm,angle=0]{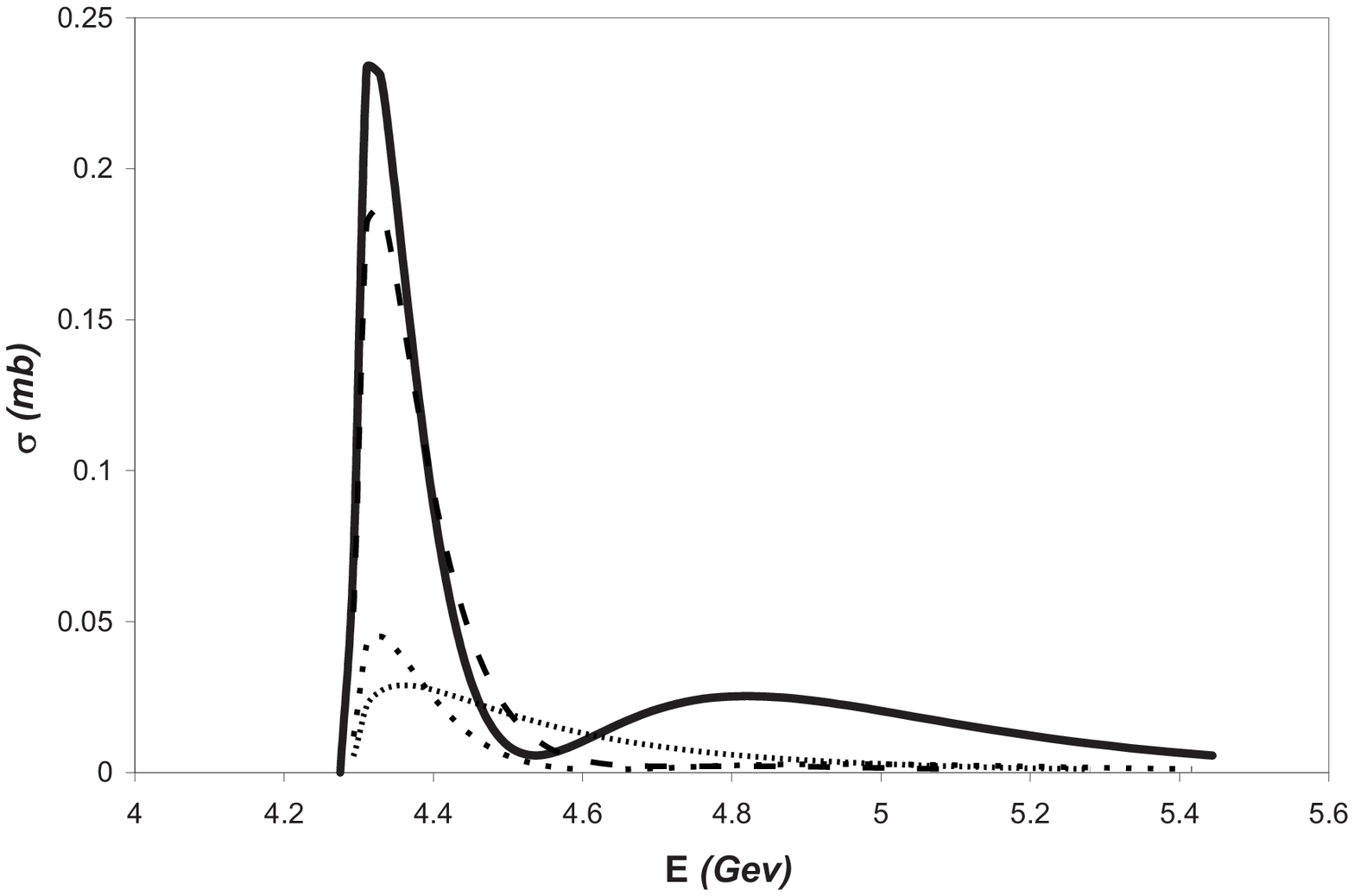}
\caption{$J/\psi p \to \bar D^{0*}\Sigma_c^+$ (left) $\eta_c p \to D^- \Sigma_c^{++}$ (right).
Curves are: total cross section (solid), hyperfine (dotted), linear (dashed), Coulomb (small dash).}
\label{xsec2Fig}
\end{figure}

The cross sections we find using this approach typically rise rapidly 
at threshold, and then are strongly damped above a hadron momentum
scale set by the hadronic wavefunctions. Since the hyperfine, Coulomb, 
and linear confinement contributions vary in sign within channels and 
have very different momentum dependences, secondary peaks can appear 
(see Figs.\ref{xsecFig},\ref{xsec2Fig}), although these tend to be much weaker 
than the near-threshold maxima. In three instances this pattern is reversed; 
one of these is shown in the right panel of Fig.\ref{xsecFig}.

We summarize all 23 independent cross sections in 
Tables~\ref{xsec1Tab}-\ref{xsec3Tab}. 
The columns in these tables 
specify the energies and values of the near-threshold maxima (point 1), 
the subsequent minimum (point 2), and the secondary maximum (point 3), 
if it is significant.

\begin{table}[h]
\caption{$\eta_c p$ Cross Sections.}
\begin{tabular}{l|lll}
    Final State     &  $(E (GeV), \sigma (mb))_1$  &  $(E (GeV), \sigma (mb))_2$  &  $(E (GeV), \sigma (mb))_3$ \\
\hline
        $\bar D^0 \Lambda_c^+$   &  (4.196 ,  1.190) & & \\
        $\bar D^0 \Sigma_c^+$    &  (4.355 ,  0.221) & & \\
        $D^- \Sigma_c^{++}$   &  (4.360 ,  0.432) & & \\
        $\bar D^{0*} \Lambda_c^+$   &  (4.311 ,  0.233) & (4.527 ,  0.006) & (4.820 ,  0.025) \\
        $\bar D^{0*} \Sigma_c^+$    &  (4.5021 , 0.0073) & (4.8162 , 0.0001) & (5.1164 , 0.0003) \\
        $\bar D^{0*} \Sigma^+_{c\, 3/2}$   &  (4.547 ,  0.018) & (4.702 ,  0.003) & (4.966 ,  0.008) \\
        $D^{-*} \Sigma_c^{++}$   &  (4.5050 , 0.0144) & (4.8195 , 0.0001) & (5.1199 , 0.0005) \\
        $D^{-*} \Sigma^{++}_{c\, 3/2}$      &  (4.551 ,  0.009) & (4.706 ,  0.001) & (4.971 ,  0.004)
\end{tabular}
\label{xsec1Tab}
\end{table}

\begin{table}[h]
\caption{$J/\psi p$ $(S=1/2)$ Cross Sections.}
\begin{tabular}{l|lll}
    Final State     &  $(E (GeV), \sigma (mb))_1$  &  $(E (GeV), \sigma (mb))_2$  &  $(E (GeV), \sigma (mb))_3$ \\
\hline
        $\bar D^0 \Lambda_c^+$   &  (4.184 ,  6.815) & (4.619 ,  0.010) & (4.884 ,  0.022) \\
        $\bar D^0 \Sigma_c^+$    &  (4.350 ,  0.126) & & \\
        $D^- \Sigma_c^{++}$   &  (4.355 ,  0.246) & & \\
        $\bar D^{0*} \Lambda_c^+$   &  (4.308 ,  0.120) & (4.425 ,  0.001) & (4.703 ,  0.079) \\
        $\bar D^{0*} \Sigma_c^+$    &  (4.498 ,  0.180) & & \\
        $\bar D^{0*} \Sigma^+_{c\, 3/2}$   &  (4.545 ,  0.009) & (4.668 ,  0.002) & (4.891 ,  0.005) \\
        $D^{-*} \Sigma_c^{++}$   &  (4.501 ,  0.354) & & \\
        $D^{-*} \Sigma^{++}_{c\, 3/2}$      &  (4.549 ,  0.017) & (4.672 ,  0.004) & (4.896 ,  0.009)
\end{tabular}
\label{xsec2Tab}
\end{table}

\begin{table}[h]
\caption{$J/\psi p$ $(S=3/2)$ Cross Sections.}
\begin{tabular}{l|lll}
    Final State     &  $(E (GeV), \sigma (mb))_1$  &  $(E (GeV), \sigma (mb))_2$  &  $(E (GeV), \sigma (mb))_3$ \\
\hline
        $\bar D^0 \Sigma^+_{c\, 3/2}$   &  (4.418 ,  0.276) & (4.873 ,  0.000) & (5.165 ,  0.001) \\
        $D^- \Sigma^{++}_{c\, 3/2}$      &  (4.424 ,  0.536) & (4.881 ,  0.001) & (5.173 ,  0.002) \\
        $\bar D^{0*} \Lambda_c^+$   &  (4.324 ,  1.712) & (4.639 ,  0.021) & (4.791 ,  0.027) \\
        $\bar D^{0*} \Sigma_c^+$    &  (4.498 ,  0.054) & & \\
        $\bar D^{0*} \Sigma^+_{c\, 3/2}$   &  (4.545 ,  0.042) & (4.733 ,  0.002) & (4.986 ,  0.006) \\
        $D^{-*} \Sigma_c^{++}$   &  (4.501 ,  0.106) & & \\
        $D^{-*} \Sigma^{++}_{c\, 3/2}$      &  (4.549 ,  0.082) & (4.737 ,  0.004) & (4.991 ,  0.012)
\end{tabular}
\label{xsec3Tab}
\end{table}

\section{Discussion and Conclusions}

The most important process is $J/\psi p \to \bar D^0 \Lambda_c^+$ scattering, with a peak cross section
of approximately 7 mb (Fig. \ref{xsecFig}). Next in strength is $J/\psi p \to
D^{0*} \Lambda_c^+$, with a maximum near 1.7 mb.  
These results disagree with meson exchange models, for example
Liu, Ko, and Lin\cite{mex2} find a larger cross section for $D^* \Lambda_c$ scattering than $D\Lambda_c$ while Sibirtsev, Tsushima, and Thomas\cite{mex2} find roughly comparable cross sections. 

\begin{figure}[h]
\includegraphics[width=7cm,angle=0]{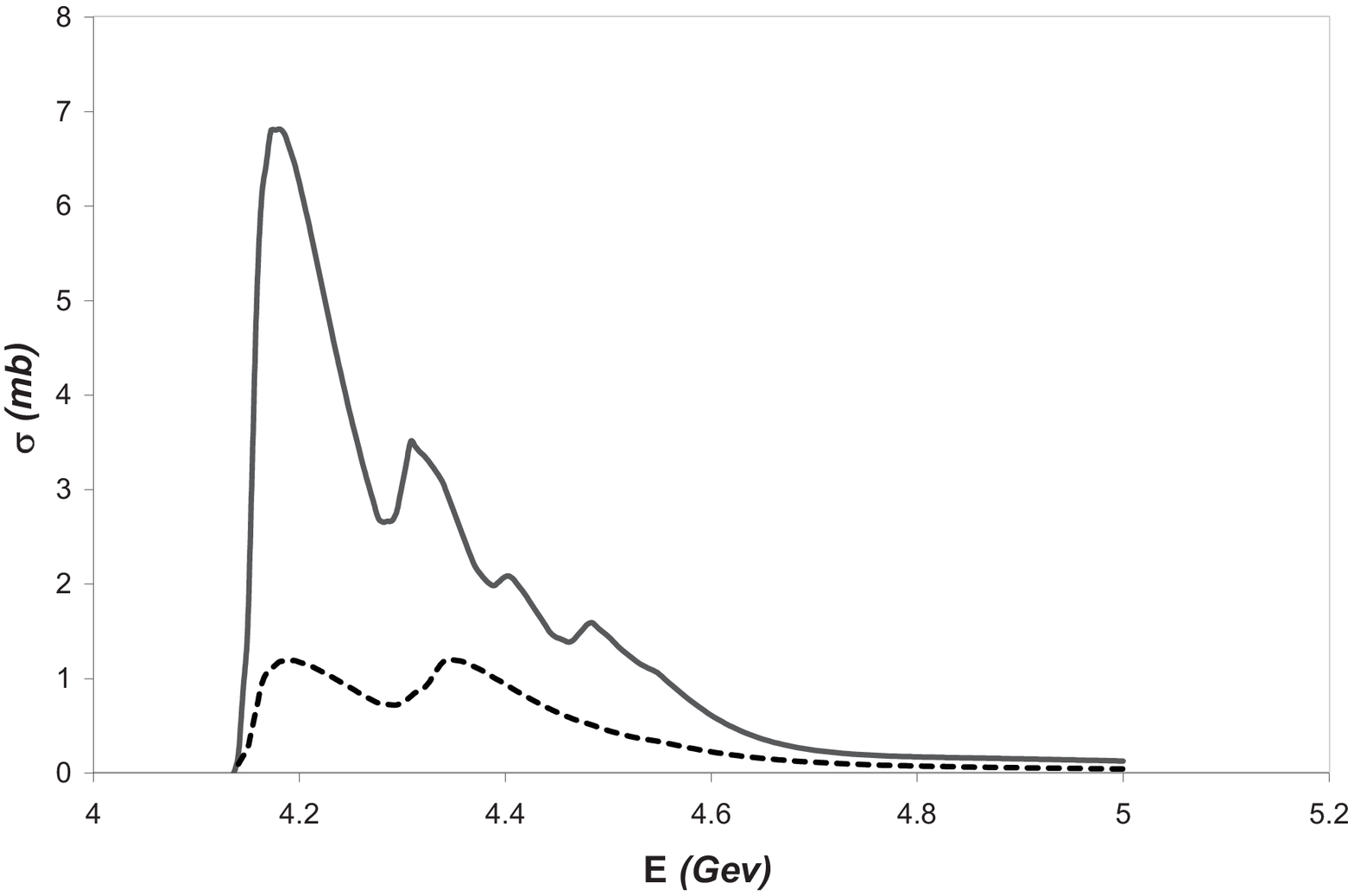}
\caption{Total Charmonium  Dissociation Cross Sections. $J/\psi N$ (solid); $\eta_c N$ (dashed).}
\label{totFig}
\end{figure}

The total dissociation cross section in shown in Fig. \ref{totFig}. It is clear that the quark model description of charm
dissociation is not in agreement with the widely held view that the absorption cross section should be constant. In general
cross sections increase with phase space and then are strongly damped due to the finite size of hadronic wavefunctions. Although
this effect can be overcome by the opening of many channels, it is evident that this does not occur up to center of mass
energies of 5 GeV or more.  This observation is in disagreement with typical meson exchange models, which tend to find 
slowly increasing total cross sections. Presumably this behaviour is due to (approximately) cancelling momentum dependence
between vertices with derivative coupling and monopole form factors. While the momentum dependence of quark model vertices must 
agree
with their hadronic analogues (see, for example, Ref. \cite{clark}), quark model form factors fall much more rapidly in
momentum than the power law monopole form factors, providing a possible source of the discrepancy.

We remark that the quark exchange computation presented here can be regarded as the short range component of $J/\psi$
dissociation in scattering from nucleons. Long range dissociation can occur via virtual pion
emission from the nucleon giving rise to reactions such as $J/\psi N \to \bar D D^* N$. This amplitude can be related 
to that of $J/\psi \pi \to \bar D D^*$ scattering,  which is relevant to comover absorption.
This has been done, for example, by Liu, Ko, and Lin\cite{mex2} who obtain asymptotic cross sections of 
order 1 mb for each of $J/\psi N \to \bar D D^* N$, $\bar D^* D N$, and $\bar D D N$. Of course, their results depend
on theoretical $J/\psi \pi$ scattering cross sections as computed in the meson exchange formalism.
Unfortunately, these are strongly dependent on form factors; for example $J/\psi \pi \to \bar D D^*$ 
ranges from 20 mb at large $\sqrt{s}$ with no form factor to 4 mb with a form factor cutoff of 1 GeV.
Alternatively, quark model computations of the same reaction give maximum cross sections of order 1/2 mb 
just above threshold\cite{wsb}, in keeping with phenomenological estimates of comover absorption\cite{ghq,cfk}.
From these considerations we conclude that the long range contribution to charmonium dissociation by nucleons is small
(less that 1 mb).

Comparison with experiment is difficult at present. As mentioned in the Introduction, $J/\psi$ photoproduction and
$p+A$ collisions yield estimates of 3.5 mb and 7 mb respectively. However these estimates are made assuming constant
cross sections, and it appears unlikely that this is accurate. Nevertheless, the average near-threshold strength of
our total cross section seems in accord with these phenomenological estimates.

Finally, 
the strong near-threshold
behaviour seen here implies that it would be worthwhile to revisit hydrodynamic simulations of charmonium suppression
in heavy ion collisions.
In this regard, we note that the current computation deals with free charmonium scattering, rather than the ``pre''-charmonium
expected in RHIC collisions. Furthermore, substantial density dependence may be present\cite{awt} and this
should be taken into account when applying these predictions to a nuclear environment.

\acknowledgments
We are grateful to Che-Ming Ko and Berndt Mueller for communications on this research.
This work was supported by the U.S. Department of Energy under contract DE-FG02-00ER41135 and
the U.S. National Science Foundation under grant NSF-PHY-244668 at the University of Pittsburgh,
the U.S. National Science
Foundation through grant NSF-PHY-0244786 at the University of Tennessee,
and the U.S. Department of Energy under contract DE-AC05-00OR22725 at
Oak Ridge National Laboratory.

\appendix

\section{Hadronic Wavefunctions}
\label{app1}

Total baryon wavefunctions were constructed as follows:

\begin{equation}
N = {\cal C} \Xi_N \Phi_B \chi^\lambda,
\end{equation}

\begin{equation}
\Lambda_c = {\cal C} \Xi_\Lambda \Phi_D \chi^\rho,
\end{equation}

\begin{equation}
\Sigma_c = {\cal C} \Xi_\Sigma \Phi_D \chi^\lambda,
\end{equation}

\begin{equation}
\Sigma_{c\, 3/2} = {\cal C} \Xi_\Sigma \Phi_D \chi^{3/2}
\end{equation}
where

\begin{equation}
\Xi_\Lambda = \frac{1}{\sqrt{2}}c( ud - du),
\end{equation}
\begin{eqnarray}
\Xi_{\Sigma^{++}} &=& cuu, \\
\Xi_{\Sigma^+} &=& \frac{1}{\sqrt{2}}c(ud + du),  \\
\Xi_{\Sigma^0} &=& cdd.
\end{eqnarray}

The baryon color wavefunction is given by
\begin{equation}
{\cal C} = \frac{1}{\sqrt{6}} \epsilon_{abc}
\end{equation}
where $\epsilon$ is the rank three antisymmetric Cartesian tensor in color indices.

Baryon spin wavefunctions are (only top states are given)

\begin{equation}
\chi^\lambda_{\frac{1}{2}\frac{1}{2}} = -\frac{1}{\sqrt{6}}( |++-\rangle + |+-+\rangle - 2|-++\rangle),
\end{equation}

\begin{equation}
\chi^\rho_{\frac{1}{2}\frac{1}{2}} = \frac{1}{\sqrt{2}}(|++-\rangle - |+-+\rangle ),
\end{equation}
and
\begin{equation}
\chi^{3/2}_{\frac{3}{2}\frac{3}{2}} = |+++\rangle .
\end{equation}

The spatial wavefunctions are 
\begin{align}
    \Phi_A &= 
\frac{1}{\beta_A^{3/2}\pi^{3/4}}
\,\textrm{exp}\left(-\frac{p_{rel}^2}{2 \beta_A^2}\right) \\
    \Phi_B &= 
\frac{3^{3/4}}{\pi^{3/2}\alpha^3}
\,\textrm{exp}\left(-\frac{p_{\rho}^2
        + p_{\lambda}^2}{2\alpha_{\rho}^2}\right) \\
    \Phi_C &= 
\frac{1}{\beta_C^{3/2}\pi^{3/4}}
\,\textrm{exp}\left(-\frac{p_{rel}^2}{2 \beta_C^2}\right) \\
    \Phi_D &= 
\frac{3^{3/4}}{\pi^{3/2}\alpha_\rho^{3/2}\alpha_{\lambda}^{3/2}}
\,\textrm{exp}\left(-\frac{p_{\rho}^2}{2 \alpha_{\rho}^2}
        - \frac{p_{\lambda}^2}{2\alpha_{\lambda}^2}\right)
\end{align}
with
\begin{align}
    p_{rel} &= \frac {m_2 p_1 - m_1 p_2}{m_1+m_2}, \\
    p_{\rho} &= \sqrt{\frac{1}{2}} \frac{\left( m_3+ 2m_5 \right) p_4
        - \left( m_3+2m_4 \right) p_5+ \left(m_5-m_4 \right) p_3 }
        {m_4+m_5+m_3}, \\
    p_{\lambda} &= \sqrt{\frac{3}{2}} \frac { m_3 \left( p_4 + p_5 \right)
        - \left( m_4+m_5 \right) p_3 }{m_4+m_5+m_3}.
\end{align}

\begin{figure}[h]
\includegraphics[width=8cm,angle=0]{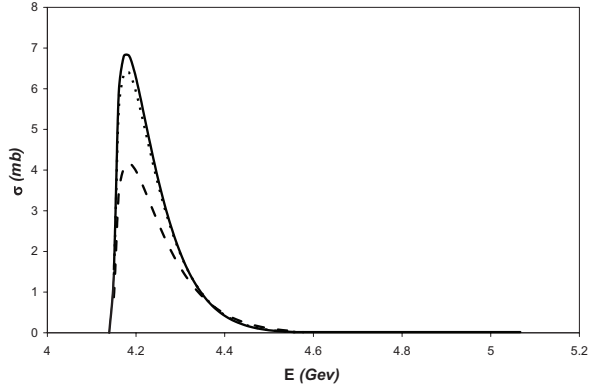}
\caption{$\sigma(\psi p \to \bar D^0 \Lambda_c)$. Single $J/\psi$ SHO (dashed), six Gaussians (dotted), ten Gaussians (solid).}
\label{wfFig}
\end{figure}

SHO meson and baryon wavefunctions can be expected to be reasonably accurate for hadrons composed
of light quarks or heavy-light mesons. However, they are not reliable for Coulombic systems such
as charmonium. Thus in our computations the nominal charmonium wavefunction ($\Phi_A$ above) has been 
replaced with
a sum over Gaussians that accurately replicates the numerically obtained wavefunction.
The subsequent change in the cross section is illustrated in Fig. \ref{wfFig}, where the importance
of an accurate charmonium wavefunction near scattering threshold is evident.

All model parameters have been fixed by meson and baryon spectroscopy. The parameters used here are listed
in Table \ref{paramTab}.

\begin{table}[h]
\caption{Model Parameters. All units are GeV except $b$ (GeV$^2$) and $\alpha_s$.}
\begin{tabular}{ccccccccccc}
$m_u$    & $m_c$     & $b$  & $\alpha_s$    & $\beta_\psi$ & $\beta_\eta$ & $\beta_D$  & $\beta_{D^*}$ & $\alpha_N$  & $\alpha_\rho$  & $\alpha_\lambda$  \\
\hline
0.33    & 1.5       & 0.162     & 0.594         & 0.67          & 0.74          & 0.44          & 0.38              & 0.3       & 0.3           & 0.45 \\
\end{tabular}
\label{paramTab}
\end{table}
Note that in an SHO (Isgur-Karl) model one has the relationships
$$\alpha_\rho = (3 k m)^{1/4}$$
and
$$\alpha_\lambda = (3 k m_\lambda)^{1/4}$$
with
$$m_\lambda = 
\frac{3 m M}{2m+M}
$$
where $M$ is the mass of quark 3.

\section{Weights}
\label{wApp}

The color matrix elements required in the evaluation of the scattering 
amplitudes we evaluate are listed in Table~\ref{ColTab} below.

\begin{table}[h]
\caption{Color Matrix Elements}
\label{ColTab}
\begin{tabular}{l|rrrrrr}
perm & 13 & 14 & 15 & 23 & 24 & 25 \\
\hline
$P_{23}$ & $-\frac{4}{9}$ & $\frac{2}{9}$ & $\frac{2}{9}$ & $\frac{4}{9}$ & $-\frac{2}{9}$ & $-\frac{2}{9}$ \\
$P_{23}P_{34}$ & $-\frac{2}{9}$ & $\frac{4}{9}$ & $-\frac{2}{9}$ & $\frac{2}{9}$ & $-\frac{4}{9}$ & $\frac{2}{9}$ \\
\end{tabular}
\end{table}

With the spatial symmetries specified in Sect. \ref{formSect}, weights are computed as follows

\be
w_1 = \langle P_{23} \Xi_C\Xi_D|\Xi_A\Xi_B\rangle\, \langle P_{23} \chi_C\chi_D|\chi_A\chi_B\rangle\,
 \langle P_{23} {\cal C}_C {\cal C}_D|T_2\cdot T_3|{\cal C}_A {\cal C}_B\rangle 
\ee
and
\be
w_2 = \langle P_{23} \Xi_C\Xi_D|\Xi_A\Xi_B\rangle\, \langle P_{23} \chi_C\chi_D|\chi_A\chi_B\rangle\,
 \langle P_{23} {\cal C}_C {\cal C}_D|T_2\cdot T_4 + T_2\cdot T_5|{\cal C}_A {\cal C}_B\rangle .
\ee
Similar expressions hold for $w_3$ and $w_4$. Here we consider the case of $P_{23}$ scattering through
a central potential. Substituting the color matrix elements of Table \ref{ColTab} then yields

\be
\vec w = \frac{4}{9} \langle P_{23} \Xi_C\Xi_D|\Xi_A\Xi_B\rangle\, \langle P_{23} \chi_C\chi_D|\chi_A\chi_B\rangle
\, (1, -1, -1, 1).
\ee
The prefactor in this expression is reported in the column labelled `Coulomb and linear' in the 
tables below. The spin-dependence of the hyperfine interaction does not permit simplification, hence
the entire weight vectors are specified.

\begin{table}[h]
\caption{$J/\psi p$, $S=1/2$ Weights.}
\begin{tabular}{l|rr}
Final State & hyperfine & Coulomb and linear \\
\hline
$\bar D^0\Lambda_c^+$ & $\frac{\sqrt{2}}{12}\left[ -1, 0, 3, 0\right]$ & $\frac{\sqrt{2}}{3}$ \\
$\bar D^0\Sigma_c^+$ & $\frac{\sqrt{6}}{108}\left[ 1, -2, -3, -2\right]$ & $-\frac{\sqrt{6}}{27}$ \\
$\bar D^{0*}\Lambda_c^+$ & $\frac{\sqrt{6}}{36}\left[ 5, 0 , 1, 0\right]$ & $-\frac{\sqrt{6}}{9}$ \\
$\bar D^{0*}\Sigma_c^+$ & $\frac{\sqrt{2}}{108}\left[ 7, 10 , -5, -6\right]$ & $\frac{5\sqrt{2}}{27}$ \\
$D^{-}\Sigma_c^{++}$ & $\frac{\sqrt{3}}{54}\left[ -1, 2 , 3, 2\right]$ & $\frac{2\sqrt{3}}{27}$ \\
$D^{-*}\Sigma_c^{++}$ & $\frac{1}{54}\left[ -7, -10 , 5, 6\right]$ & $-\frac{10}{27}$ \\
$\bar D^{0*}\Sigma^+_{c\,3/2}$ & $\frac{1}{27}\left[ 1, 1 , 1, 3\right]$ & $\frac{4}{27}$ \\
$D^{-*}\Sigma^{++}_{c\,3/2}$ & $-\frac{\sqrt{2}}{27}\left[ 1, 1 , 1, 3\right]$ & $\frac{4\sqrt{2}}{27}$ \\
\end{tabular}
\end{table}

\begin{table}[h]
\caption{$J/\psi p$, $S=3/2$ Weights.}
\begin{tabular}{l|rr}
Final State & hyperfine & Coulomb and linear \\
\hline
$\bar D^{0*}\Lambda_c^+$ & $\frac{\sqrt{6}}{18} \left[ 1, 0, -1, 0\right]$ & $\frac{2\sqrt{6}}{9}$ \\
$\bar D^{0*}\Sigma_c^+$ & $\frac{\sqrt{2}}{54} \left[ 5, 2, -1, 0\right]$ & $\frac{2\sqrt{2}}{27}$ \\
$D^{-*}\Sigma_c^{++}$ & $-\frac{1}{27} \left[ 5, 2, -1, 0\right]$ & $-\frac{4}{27}$ \\
$\bar D^{0*}\Sigma^{+}_{c\,3/2}$ & $-\frac{\sqrt{10}}{54} \left[ 1, 1, 1, 0\right]$ & $\frac{2\sqrt{10}}{27}$ \\
$D^{-*}\Sigma^{++}_{c\,3/2}$ & $-\frac{\sqrt{5}}{27} \left[ 1, 1, 1, 0\right]$ & $-\frac{4\sqrt{5}}{27}$ \\
$\bar D^{0}\Sigma^{+}_{c\,3/2}$ & $-\frac{\sqrt{6}}{54} \left[ 1, 1, -3, -2\right]$ & $\frac{2\sqrt{6}}{27}$ \\
$D^{-}\Sigma^{++}_{c\,3/2}$ & $\frac{\sqrt{3}}{27} \left[ 1, 1, -3, -2\right]$ & $-\frac{4\sqrt{3}}{27}$ \\
\end{tabular}
\end{table}

\begin{table}[h]
\caption{$\eta_c p$ Weights.}
\begin{tabular}{l|rr}
Final State & hyperfine & Coulomb and linear \\
\hline
$\bar D^{0}\Lambda_c^{+}$ & $\frac{\sqrt{6}}{12}\left[ 1, 0, 1, 0\right]$ & $\frac{\sqrt{6}}{9}$ \\
$\bar D^{0}\Sigma_c^{+}$ & $\frac{\sqrt{2}}{36}\left[ 3, 2, 3, 2\right]$ & $\frac{\sqrt{2}}{9}$ \\
$\bar D^{0*}\Lambda_c^{+}$ & $-\frac{\sqrt{2}}{12}\left[ 1, 0, 1, 0\right]$ & $\frac{\sqrt{2}}{3}$ \\
$\bar D^{0*}\Sigma_c^{+}$ & $\frac{\sqrt{6}}{108}\left[ 1, -2, 1, -2\right]$ & $-\frac{\sqrt{6}}{27}$ \\
$D^{-}\Sigma_c^{++}$ & $-\frac{1}{18}\left[ 3, 2, 3, 2\right]$ & $-\frac{2}{9}$ \\
$D^{-*}\Sigma_c^{++}$ & $-\frac{\sqrt{3}}{54}\left[ 1, -2, 1, -2\right]$ & $\frac{2\sqrt{3}}{27}$ \\
$\bar D^{0*}\Sigma^{+}_{c\,3/2}$ & $\frac{\sqrt{12}}{54}\left[ 1, 1, 1, 1\right]$ & $-\frac{2\sqrt{12}}{27}$ \\
$D^{-*}\Sigma^{++}_{c\,3/2}$ & $-\frac{\sqrt{6}}{27}\left[ 1, 1, 1, 1\right]$ & $\frac{4\sqrt{6}}{27}$ \\
\end{tabular}
\end{table}


\end{document}